%% file: main.tex
\RequirePackage{amsmath}
\documentclass[runningheads]{llncs}

\usepackage{algpseudocode}
\usepackage{algorithm}
\algrenewcommand\algorithmicindent{0.6em}%
\usepackage{amssymb}
\usepackage{caption}
\usepackage{graphicx}
\usepackage[english]{babel}
\usepackage[backend=biber,sorting=nyt,bibencoding=latin1,abbreviate=false,mincrossrefs=3,style=numeric,maxnames=30]{biblatex}
\usepackage[inline]{enumitem}
\usepackage{hyperref}
\usepackage{subcaption}
\usepackage{todonotes}

\newcommand{\pe}[0]{p_e}
\newcommand{\peq}[0]{p_{eq}}
\newcommand{\ps}[0]{p_s}
\newcommand{\psub}[0]{p_{sub}}

\newcommand{\cl}[3]{\mathcal{#1}\substack{#2\\#3}}

\newcommand{\cli}[3]{\overline{\mathcal{#1}\substack{#2\\#3}}}
\newcommand{\clit}[3]{\overline{\mathcal{#1}^+\substack{#2\\#3}}}

\DeclareMathOperator*{\argmax}{argmax}

\begin{document}

\title{Observing LOD using Equivalent Set Graphs: \\it is mostly flat and sparsely linked}

\titlerunning{Observing the LOD Cloud
using Equivalent Set Graphs}

\author{Luigi Asprino\inst{1,2} \and Wouter Beek\inst{3} \and Paolo Ciancarini \inst{2} \and Frank van Harmelen \inst{3} \and Valentina Presutti \inst{1}}

\authorrunning{Asprino et al.}

\institute{STLab, ISTC-CNR, Rome, Italy \\
\email{luigi.asprino@istc.cnr.it valentina.presutti@cnr.it} 
\and
University of Bologna, Bologna, Italy\\
\email{ paolo.ciancarini@unibo.it}  \and
Dept. of Computer Science, VU University Amsterdam, NL\\
\email{\{w.g.j.beek, frank.van.harmelen\}@vu.nl}}

\maketitle          

\begin{abstract}
This paper presents an empirical study aiming at understanding the modeling style and the overall semantic structure of Linked Open Data. We observe how classes, properties and individuals are used in practice. We also investigate how hierarchies of concepts are structured, and how much they are linked. In addition to discussing the results, this paper contributes 
\begin{enumerate*}[label=\textit{(\roman*)}]

\item a conceptual framework, including a set of metrics, which generalises over the observable constructs;

\item an open source implementation that facilitates its application to other Linked Data knowledge graphs.
\end{enumerate*}


\end{abstract}

\section{Analysing the modeling structure and style of LOD}
\label{sec:intro}
\input{src/intro.tex}

\section{Related Work}
\input{src/related.tex}
\label{sec:related}
%

\section{Approach}
\label{sec:approach}

\subsection{{Input source}} 
\label{src:input}
\input{src/input.tex}

This section outlines our approach for performing large-scale semantic analyses of the LOD KG. We start out by introducing the new notion of Equivalence Set Graph (ESG) (Section~\ref{sec:introducing-equivalence-set-graphs}). Once Equivalence Set Graphs have been informally introduced, the corresponding formal definitions are given in Section~\ref{sec:formalizing-equivalence-set-graphs}. Finally, the metrics that will be measured using the ESGs are defined in Section~\ref{sec:metrics}.

\subsection{Introducing Equivalence Set Graphs}
\label{sec:introducing-equivalence-set-graphs}
\input{src/graph.tex}

\subsection{Metrics}
\label{sec:metrics}
\input{src/metrics.tex}

\section{Computing Equivalence Set Graphs}
\label{sec:algorithm}
\input{src/algorithm.tex}

\section{Results}
\label{sec:stats}
\input{src/lodGraph.tex}

\section{Discussion}
\label{sec:discussion}
\input{src/discussion.tex}


\printbibliography
\end{document}

%% file: src/intro.tex
The interlinked collection of Linked Open Data (LOD) datasets forms the largest publicly accessible Knowledge Graph (KG) that is available on the Web today.\footnote{This paper uses the following RDF prefix declarations for brevity, and uses the empty prefix (\texttt{:}) to denote an arbitrary example namespace.
\begin{itemize}
  \item \texttt{dbo}: \url{http://dbpedia.org/ontology/}
  \item \texttt{dul}: \url{http://www.ontologydesignpatterns.org/ont/dul/DUL.owl\#}
  \item \texttt{foaf}: \url{http://xmlns.com/foaf/0.1/}
  \item \texttt{org}: \url{http://www.w3.org/ns/org\#}
  \item \texttt{rdfs}: \url{http://www.w3.org/2000/01/rdf-schema\#}
  \item \texttt{owl}: \url{http://www.w3.org/2002/07/owl\#}
\end{itemize}} 
LOD distinguishes itself from most other forms of open data in that it has a formal semantics. Various studies have analysed different aspects of the formal semantics of LOD.  However, existing analyses have often been based on relatively small samples of the ever evolving LOD KG.  Moreover, it is not always clear how representative the chosen samples are.  This is especially the case when observations are based on one dataset (e.g., DBpedia), or on a small number of datasets that are drawn from the much larger LOD Cloud.

This paper presents observations that have been conducted across (a very large subset of) the LOD KG.  As such, this paper is not about the design of individual ontologies, rather, it is about observing \emph{the design of the globally shared Linked Open Data ontology}.  Specifically, this paper focuses on the globally shared hierarchies of classes and properties, together with their usage in instance data.  This paper provides new insights about
\begin{enumerate*}[label=\textit{(\roman*)}]
    \item the number of concepts defined in the LOD KG,
    \item the shape of ontological hierarchies,
    \item the extent in which recommended practices for ontology alignment are followed, and 
    \item whether classes and properties are instantiated in a homogeneous way.
\end{enumerate*}

In order to conduct large-scale semantic analyses, it is necessary to calculate the deductive closure of very large hierarchical structures.  Unfortunately, contemporary reasoners cannot be applied at this scale, unless they rely on expensive hardware such as a multi-node in-memory cluster.  In order to handle this type of large-scale semantic analysis on commodity hardware such as regular laptops, we introduce the formal notion of an \emph{Equivalence Set Graph}.  With this notion we are able to implement efficient algorithms to build the large hierarchical structures that we need for our study.

We use the formalization and implementation presented in this paper to compute two (very large) Equivalence Set Graphs: one for classes and one for properties.  By querying them, we are able to quantify various aspects of formal semantics at the scale of the LOD KG.  Our observations show that there is a lack of explicit links (alignment) between ontological entities and that there is a significant number of concepts with empty extension.  Furthermore, property hierarchies are observed to be mainly flat, while class hierarchies have varying depth degree, although most of them are flat too.

This paper makes the following contributions:

\begin{enumerate}
  
  \item A new formal concept (Equivalence Set Graph) that allows us to specify compressed views of a LOD KG (presented in Section~\ref{sec:introducing-equivalence-set-graphs}).
  
  \item An implementation of efficient algorithms that allow Equivalence Set Graphs to be calculated on commodity hardware (cf. Section~\ref{sec:algorithm}).
  
  \item A detailed analysis of how classes and properties are used at the level of the whole LOD KG, using the formalization and implementation of Equivalence Set Graphs.
  
\end{enumerate}

The remaining of this paper is organized as follows: Section~\ref{sec:related} summarizes related work. The approach is presented in Section~\ref{sec:approach}. 
Section~\ref{sec:algorithm} describes the algorithm for computing an Equivalence Set Graph form a RDF dataset.
Section~\ref{sec:metrics} defines a set of metrics that are measured in Section \ref{sec:stats}. Section~\ref{sec:discussion} discusses the observed values and concludes.

%% file: src/related.tex
Although large-scale analyses of LOD have been performed since the early years of the Semantic Web, we could not find previous work directly comparable with ours.
The closest we found are not recent and performed on a much smaller scale.
In 2004, Gil and Garc{\'i}a~\cite{shortGill2004} showed that the Semantic Web (at that time consisting of 1.3 million triples distributed over 282 datasets) behaves as a Complex System: the average path length between nodes is short (small world property), there is a high probability that two neighbors of a node are also neighbors of one another (high clustering factor), and nodes follow a power-law degree distribution.  In 2008, similar results were reported by \cite{shortTheoharis2008} in an individual analysis of 250 schemas. These two studies focus on topological graph aspects exclusively, and do not take semantics into account.

In 2005, Ding et al.~\cite{shortDing2005} analysed the use of the Friend-of-a-Friend (FOAF) vocabulary on the Semantic Web.  They harvested 1.5 million RDF datasets, and computed a social network based on those data datasets.  They observed that the number of instances per dataset follows the Zipf distribution.

In 2006, Ding et al.~\cite{shortDing2006} analysed 1.7 million datasets, containing 300 million triples.  They reported various statistics over this data collection, such as the number of datasets per namespace, the number of triples per dataset, and the number of class- and property-denoting terms.  The semantic observation in this study is limited since no deduction was applied.

In 2006, a survey by Wang et al.~\cite{shortWang2006} aimed at assessing the use of OWL and RDF schema vocabularies in 1,300 ontologies harvested from the Web. This study reported statistics such as the number of classes, properties, and instances of these ontologies. Our study provides both an updated view on these statistics, and a much larger scale of the observation (we analysed ontological entities defined in $\sim$650k datasets crawled by LOD-a-lot~\cite{shortFernandez2017}).

Several studies~\cite{shortBeek2018,shortDing2010,shortHalpin2010} analysed common issues with the use of \texttt{owl:sameAs} in practice.  
Mallea et al.~\cite{shortMallea2011} showed that blank nodes, although discouraged by guidelines, are prevalent on the Semantic Web. 
Recent studies~\cite{shortPaulheim2015} experimented on analysing the coherence of large LOD datasets, such as DBpedia, by leveraging foundational ontologies. Observations on the presence of foundational distinctions in LOD has been studied in~\cite{shortAsprino2018a}.  

These studies have a similar goal as ours: to answer the question how knowledge representation is used in practice in the Semantic Web, although the focus may partially overlap. We generalise over all equivalence (or identity) constructs instead of focusing on one specific, we observe the overall design of LOD ontologies, analysing a very large subject of it, we take semantics into account by analysing the asserted as well as the inferred data.

%% file: src/input.tex
Ideally, our input is the whole LOD Cloud, which is (a common metonymy for identifying) a very large and distributed Knowledge Graph.  
The two largest available crawls of LOD available today are WebDataCommons and LOD-a-lot. 

WebDataCommons\footnote{See \url{http://webdatacommons.org}}~\cite{shortMeusel2014} consists of $\sim$31B triples that have been extracted from the CommonCrawl datasets (November 2018 version).  Since its focus is mostly on RDFa, microdata, and microformats, WebDataCommons contains a very large number of relatively small graph components that use the Schema.org\footnote{See \url{https://schema.org}} vocabulary.

LOD-a-lot\footnote{See \url{http://lod-a-lot.lod.labs.vu.nl}}~\cite{shortFernandez2017} contains $\sim$28B unique triples that are the result of merging the graphs that have been crawled by LOD Laundromat~\cite{shortBeek2014} into one single graph. The LOD Laundromat crawl is based on data dumps that are published as part of the LOD Cloud, hence it contains relatively large graphs that are highly interlinked. The LOD-a-lot datadump is more likely to contain RDFS and OWL annotations than WebDataCommons. Since this study focuses on the \emph{semantics} of Linked Open Data, it uses the LOD-a-lot datadump.

LOD-a-lot only contains explicit assertions, i.e., triples that have been literally published by some data owner.  This means that the implicit assertions, i.e., triples that can be derived from explicit assertions and/or other implicit assertions, are not part of it and must be calculated by a reasoner.  Unfortunately, contemporary reasoners are unable to compute the semantic closure over 28B triples.  Advanced alternatives for large-scale reasoning, such as the use of clustering computing techniques (e.g.,~\cite{shortUrbani2010}) require expensive resources in terms of CPU/time and memory/space.  Since we want to make running large-scale semantic analysis a frequent activity in Linked Data Science, we present a new way to perform such large-scale analyses against very low hardware cost.

%% file: src/graph.tex
An Equivalence Set Graph (ESG) is a tuple $\langle \mathcal{V}, \mathcal{E}, \peq, \psub, \pe, \ps \rangle$.  The nodes $\mathcal{V}$ of an ESG are equivalence sets of terms from the universe of discourse.  The directed edges $\mathcal{E}$ of an ESG are specialization relations between those equivalence sets.  $\peq$ is an equivalence relation that determines which equivalence sets are formed from the terms in the universe of discourse.  $\psub$ is a partial order relation that determines the specialization relation between the equivalence sets. In order to handle equivalences and specializations of $\peq$ and $\psub$ (see below for details and examples), we introduce $\pe$, an equivalence relation over properties (e.g., \texttt{owl:equivalentProperty}) that allows to retrieve all the properties that are equivalent to $\peq$ and $\psub$, and $\ps$ which is a specialization relation over properties (e.g., \texttt{rdfs:subPropertyOf}) that allows to retrieve all the properties that specialize $\peq$ and $\psub$.

The inclusion of the parameters $\peq$, $\psub$, $\pe$, and $\ps$ makes the Equivalence Set Graph a very generic concept.  By changing the equivalence relation ($\peq$), ESG can be applied to classes (\texttt{owl:equivalentClass}), properties (\texttt{owl:equivalentProperty}), or instances (\texttt{owl:sameAs}).  By changing the specialization relation ($\psub$), ESG can be applied to class hierarchies (\texttt{rdfs:subClassOf}), property hierarchies (\texttt{rdfs:subPropertyOf}), or concept hierarchies (\texttt{skos:broader}).

An Equivalence Set Graph is created starting from a given RDF Knowledge Graph. The triples in the RDF KG are referred to as its \emph{explicit} statements. The \emph{implicit} statements are those that can be inferred from the explicit statements. An ESG must be built taking into account both the explicit and the \emph{implicit} statements.  For example, if $\peq$ is \texttt{owl:equivalentClass}, then the following Triple Patterns (TP) retrieve the terms \texttt{?y} that are explicitly equivalent to a given ground term \texttt{:x}:

\begin{verbatim}{sparql}
  { :x owl:equivalentClass ?y } union { ?y owl:equivalentClass :x }
\end{verbatim}

In order to identify the terms that are \textit{implicitly} equivalent to \texttt{:x}, we also have to take into account the following:
\begin{enumerate}

  \item The closure of the equivalence predicate (reflexive, symmetric, transitive).

  \item Equivalences (w.r.t. $\pe$) and/or specializations (w.r.t. $\ps$) of the equivalence predicate ($\peq$).  E.g., the equivalence between \texttt{:x} and \texttt{:y} is asserted with the \texttt{:sameClass} predicate, which is equivalent to \texttt{owl:equivalentClass}):

  \begin{verbatim}{turtle}
    :sameClass owl:equivalentProperty owl:equivalentClass.
    :x :sameClass :y.
  \end{verbatim}

  \item Equivalences (w.r.t. $\pe$) and/or specializations (w.r.t. $\ps$) of predicates (i.e. $\pe$ and $\ps$) for asserting equivalence or specialization relations  among properties .  E.g., the equivalence between \texttt{:x} and \texttt{:y} is asserted with the \texttt{:sameClass} predicate, which is a specialization of \texttt{owl:equivalentClass} according to \texttt{:sameProperty}, which it itself a specialization of \texttt{owl:equivalentProperty}:

  \begin{verbatim}
    :sameProperty rdfs:subPropertyOf owl:equivalentProperty.
    :sameClass :sameProperty owl:equivalentClass.
    :x :sameClass :y.
  \end{verbatim}
\end{enumerate}

The same distinction between explicit and implicit statements can be made with respect to the specialization relation ($\psub$).  E.g., for an Equivalence Set Graph that uses \texttt{rdfs:subClassOf} as its specialization relation, the following TP retrieves the terms \texttt{?y} that explicitly specialize a given ground term \texttt{:x}:

\begin{verbatim}
    ?y rdfs:subClassOf :x.
\end{verbatim}

In order to identify the entities that are \emph{implicit} specializations of \texttt{:x}, we must also take the following into account:

\begin{enumerate}

  \item The closure of the specialization predicate (reflexive, anti-symmetric, transitive).

  \item Equivalences (w.r.t. $\pe$) and/or specializations (w.r.t. $\ps$) of the specialization predicate ($\psub$).  E.g, \texttt{:y} is a specialization of \texttt{:x} according to the \texttt{:subClass} property, which is itself a specialization of the \texttt{rdfs:subClassOf} predicate:

  \begin{verbatim}
    :subClass rdfs:subPropertyOf rdfs:subClassOf.
    :y :subClass :x.
  \end{verbatim}
    
  \item Equivalences (w.r.t. $\pe$) and/or specializations  (w.r.t. $\ps$) of  predicates (i.e. $\pe$ and $\ps$) for asserting equivalence or specialization relations among properties:

  \begin{verbatim}
    :subProperty rdfs:subPropertyOf rdfs:subPropertyOf.
    :subClass :subProperty rdfs:subClassOf.
    :y :subClass :x.
  \end{verbatim}

\end{enumerate}
Although there exist alternative ways  for asserting an equivalence (specialization) relation between two entities $e_1$ and $e_2$  (e.g., $e_1 = e_2 \sqcap \exists p.\top$ implies $e_1 \sqsubseteq e_2$), we focused on the most explicit ones, namely, those in which $e_1$ and $e_2$ are connected by a path having as edges $\peq$ ($\psub$) or properties that are equivalent or subsumed by $\peq$ (called \textit{Closure Path} cf. Definition~\ref{def:closure-path}).
We argue that for statistical observations explicit assertions provide acceptable approximations of the overall picture.

\begin{figure}[t]
  \centering
  \begin{subfigure}[b]{0.53\textwidth}
    \includegraphics[width=\textwidth]{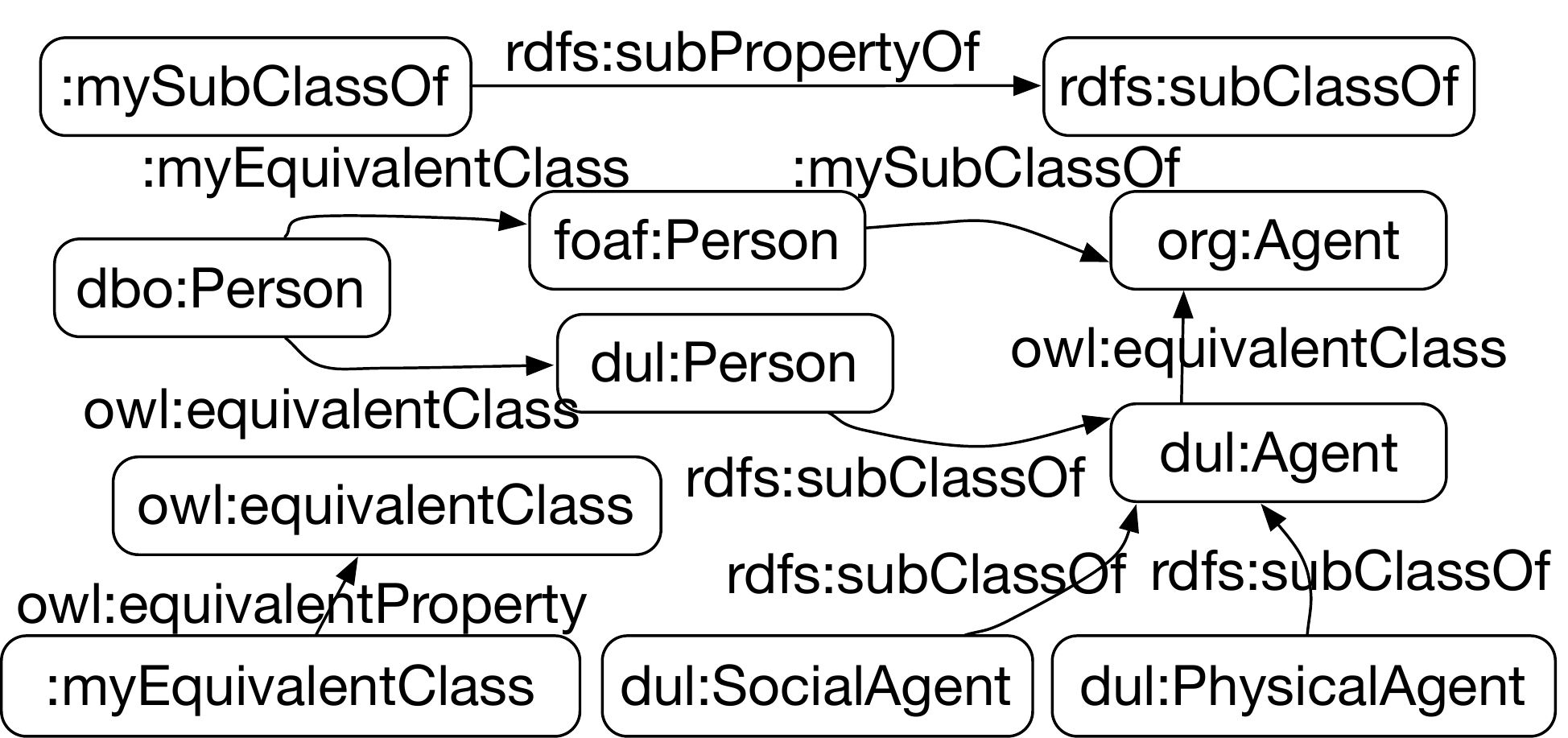}
    \caption{RDF Knowledge Graph}
    \label{fig:input}
  \end{subfigure}
  ~%
  \begin{subfigure}[b]{0.45\textwidth}
    \includegraphics[width=\textwidth]{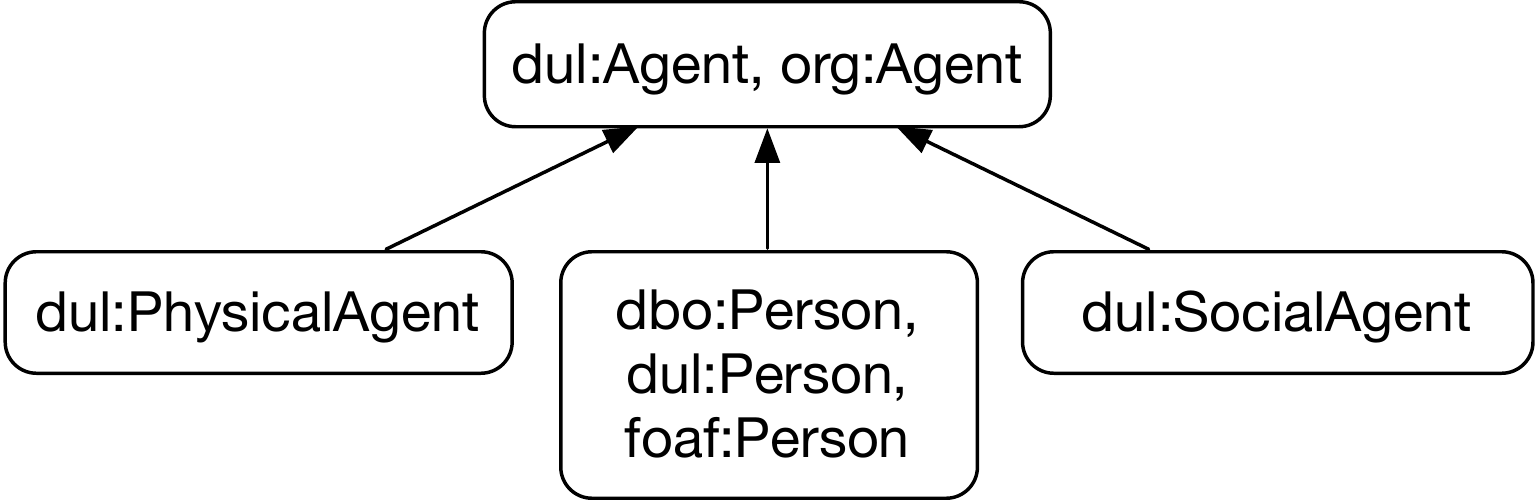}
    \caption{Equivalence Set Graph}
    \label{fig:output}
  \end{subfigure}
  \caption{An example of an RDF Knowledge Graph and its corresponding Equivalence Set Graph.}
  \label{fig:example}
\end{figure}

Figure~\ref{fig:example} shows an example of an RDF Knowledge Graph (Subfigure~\ref{fig:input}).  The equivalence predicate ($\peq$) is \texttt{owl:equivalentClass}; the specialization predicate ($\psub$) is \texttt{rdfs:subClassOf}, the property for asserting equivalences among predicates ($\pe$) is \texttt{owl:equivalentProperty}, the property for asserting specializations among predicates ($\ps$) is (\texttt{rdfs:subPropertyOf}).  The corresponding Equivalence Set Graph (Subfigure~\ref{fig:output}) contains four equivalence sets.  The top node represents the agent node, which encapsulates entities in DOLCE and W3C's Organization ontology.  Three nodes inherit from the agent node.  Two nodes contain classes that specialize \texttt{dul:Agent} in the DOLCE ontology (i.e. \texttt{dul:PhysicalAgent} and \texttt{dul:SocialAgent}).  The third node represents the person concept, which encapsulates entities in DBpedia, DOLCE, and FOAF.  The equivalence of these classes is asserted by \texttt{owl:equivalentClass} and \texttt{:myEquivalentClass}.  Since \texttt{foaf:Person} specialises \texttt{org:Agent} (using \texttt{:mySubClassOf} which specialises \texttt{rdfs:subClassOf}) and \texttt{dul:Person} specialises \texttt{dul:Agent} the ESG contains an edge between the person and the agent concept.

\subsection{Formalizing Equivalence Set Graphs}
\label{sec:formalizing-equivalence-set-graphs}

This section contains the formalization of ESGs that were informally introduced above.  An ESG must be configured with ground terms for the following  parameters:
\begin{enumerate*}[label=\textit{(\roman*)}]
  \item $\peq$: the equivalence property for the observed entities;
  \item $\psub$: the specialization property for the observed entities;
  \item $\pe$ the equivalence property for properties;
  \item $\ps$ the specialization property for properties.
\end{enumerate*}

Definition~\ref{def:deductive-closure} specifies the deductive closure over an arbitrary property $p$ with respect to $\pe$ and $\ps$. This is the set of properties that are implicitly equivalent to or subsumed by $p$. It is worth noticing that, in the general case, a deductive closure for a class of (observed) entities depends on all the four parameters: $\peq$  and $\psub$ are needed for retrieving equivalences and specializations among entities, and $\pe$ and $\ps$ are need for retrieving equivalences and specializations of $\peq$ and $\psub$. It is easy to see that when the subject of observation are properties $\peq$ and $\psub$ coincide with $\pe$ and $\ps$ respectively.

\begin{definition}[Deductive Closure of Properties]
  \label{def:deductive-closure}
  $\cli{C}{\pe,\ps}{\pe,\ps}(p)$ is the deductive closure of property $p$ with respect to $\pe$ and $\ps$.
\end{definition}

\begin{definition}[Closure Path]
  \label{def:closure-path}
  $\substack{p+\\\Longleftrightarrow}$ denotes any path, consisting of one or more occurrences of predicates from $\cli{C}{\pe,\ps}{\pe,\ps}(p)$.
\end{definition}

Once the four custom parameters have been specified, a specific Equivalence Set Graph is determined by Definitions~\ref{def:nodes} and~\ref{def:edges}.  

\begin{definition}[ESG Nodes]
  \label{def:nodes}
  Let $G$ be the graph merge~\cite{Hayes2014} of an RDF Knowledge Graph.  The set of nodes of the corresponding Equivalence Set Graph is:
  \[
    \cl{V}{\peq,\psub}{\pe,\ps} \,:=\, \{ v= \{ e_1, \ldots, e_n \} \,|\, (\forall e_i,e_j \in v)(e_i \substack{\peq+\\\Longleftrightarrow} e_j \in G)\}
  \]
\end{definition}

\begin{definition}[ESG Edges]
  \label{def:edges}
  Let $G$ be the graph merge of an RDF Knowledge Graph.  The set of edges of the corresponding Equivalence Set Graph is:
  \begin{multline*}
    \cl{E}{\peq,\psub}{\pe,\ps} \,:=\, \{ (v=\{ v_1, \ldots, v_n \}, z=\{ z_1, \ldots, z_n \}) \,\vert\\
    (\exists v_i \in v)(\exists z_j \in z)(\exists p \in  \cli{C}{\pe,\ps}{\pe,\ps}(\psub))(\langle v_i, p, z_j \rangle \in G)\}
  \end{multline*}
\end{definition}

Definitions~\ref{def:specClosure} and~\ref{def:closure} define the concept of closure.

\begin{definition}[Specialization Closure]
  \label{def:specClosure}
  Let $G$ be the graph merge of an RDF Knowledge Graph.  The specialization closure of $G$ is a function that maps an entity $e$ onto the set of entities that implicitly specialise $e$:
  \[
    \clit{H}{\peq,\psub}{\pe,\ps}(e) \,:=\, \{ e' \,|\, e' \substack{\psub+\\\Longrightarrow} e \in G \}
  \]
\end{definition}

\begin{definition}[Equivalence and Specialization Closure]
  \label{def:closure}
  Let G be a graph merge of an RDF Knowledge Graph, the equivalence and specialization closure of $G$ is a function that given an entity e returns all the entities that are either implicitly equivalent to e, or implicitly specialize e. I.e.:
  \[
    \cli{C}{\peq,\psub}{\pe,\ps}(e) \,:=\, \{ e' \,|\, (\exists v \in \cl{V}{\peq,\psub}{\pe,\ps})(e \in v \,\land\, e' \in v) \} \cup \clit{H}{\peq,\psub}{\pe,\ps}(e)
  \]
\end{definition}

%% file: src/metrics.tex
In this section we define a set of metrics that can be computed by querying Equivalence Set Graphs.

\textbf{Number of equivalence sets (ES), \textbf{Number of observed entities (OE)}, and \textbf{Ratio (R).}}
The number of equivalence sets (ES) is the number of nodes in an Equivalence Set Graph, i.e., $\vert\cl{V}{\peq,\psub}{\pe,\ps}\vert$. Equivalence sets contain equivalent entities (classes, properties or individuals). The number of observed entities (OE) is the size of the universe of discourse: i.e. $\vert \{ e \in v \,\vert\, v \in \cl{V}{\peq,\psub}{\pe,\ps} \} \vert$.  The ratio $\frac{ES}{OE}$ (R) between the number of equivalence sets and the number of entities indicates to what extent equivalence is used among the observed entities. If equivalence is rarely used, R approaches 1.0.

\textbf{Number of edges (E)}
The total number of edges is $\vert \cl{E}{\peq,\psub}{\pe,\ps} \vert$.  

\textbf{Height of Nodes.}
The height $h(v)$ of a node $v$ is defined as the length of the longest path from a leaf node until $v$. The maximum height   of an ESG is defined as $H_{max}=\argmax_{v \in V} h(v)$. 
\textit{Distribution of the height}: for n ranging from 0 to 
$H_{max}$ we compute the percentage of nodes having that height (i.e. H(n)).

\textbf{Number of Isolated Equivalent Sets (IN), Number of Top Level Equivalence Sets (TL).}
In order to observe the shape and structure of hierarchies in LOD, we compute the number Isolated Equivalent Sets (IN) in the graph, and the number of Top Level Equivalence Sets (TL).
An IES is a node without incoming or outgoing edges. A TL is a node without outgoing edges.

\textbf{Extensional Size of Observed Entities.}
Let $c$ be a class in LOD, and $t$ a property in the deductive closure of \texttt{rdf:type}. We define the extensional size of $c$ (i.e. $S(c)$) as the number of triples having $c$ as object and $t$ as predicate (i.e. $S(c) = \sum_{t\in\cli{C}{}{}} |\{\langle e, t, c\rangle|\exists e. \langle e,t,c \rangle \in G\}|$ where $\cli{C}{}{}$ is $\cli{C}{p_e,p_s}{p_e,p_s}$).
We define the extensional size of a property $p$ (i.e. $S(p)$) as the number of triples having $p$ as predicate (i.e. $S(p)=|\{\langle s, p, o\rangle |\exists p, o. \langle s, p, o\rangle \in G\}|$).

\textbf{Extensional Size of Equivalence Sets.}
We define two measures: \emph{direct extensional size} (i.e. DES) and \emph{indirect extensional size} (i.e. IES). DES is defined as the sum of the extensional size of the entities belonging to the set. The IES is its DES summed with the DES of all equivalence sets in its closure.

\noindent\textbf{Number of Blank Nodes.}
Blank nodes are anonymous RDF resource used (for example) within ontologies to define class restrictions.
We compute the number of blank nodes in LOD and we compute the above metrics both including and excluding blank nodes. 

\textbf{Number of Connected Components.}
Given a directed graph G, a strongly connected component (SCC) is a sub-graph of G where any two nodes are connected to each other by at least one path; a weakly connected component (WCC) is the undirected version of a sub-graph of G where any two nodes are connected by any path. We compute the number and the size of SCC and WCC of an ESG, to observe its distribution. Observing these values (especially on WCC) provides insights on the shape of hierarchical structures formed by the observed entities, at LOD scale.

%% file: src/algorithm.tex
In this Section we describe the algorithm for computing an equivalence set graph from a RDF dataset.
An implementation of the algorithm is available online\footnote{\url{https://w3id.org/edwin/repository}}.

\textbf{Selecting Entities to Observe.}
The first step of the procedure for computing an ESG is to select the entities to observe, from the input KG.
To this end, a set of criteria for selecting these entities can be defined. In our study we want to observe the behaviour of classes and properties, hence our criteria are the followings:
\begin{enumerate*}[label=\textit{(\roman*)}]
    \item A class is an entity  that belongs to  \texttt{rdfs:Class}.
    We assume that the property for declaring that an entity belongs to a class is \texttt{rdf:type}.
    \item A class is the subject (object) of a triple where the property has \texttt{rdfs:Class} as domain (range).
    We assume that the property for declaring the domain (range) of a property is \texttt{rdfs:domain} (\texttt{rdfs:range}).
    \item A property is the predicate of a triple.
    \item A property is an entity that belongs to \texttt{rdf:Property}.
    \item A property is the subject (object) of a triple where the property has \texttt{rdf:Property} as domain (range).
\end{enumerate*}
We defined these criteria since the object of our observation are classes and properties, but the framework can be also configured for observing other kinds of entities (e.g. individuals).

As discussed in Section~\ref{sec:introducing-equivalence-set-graphs}  we have to take into account possible equivalences and/or specializations of the ground terms, i.e. \texttt{rdf:type}, \texttt{rdfs:range}, \texttt{rdfs:domain} and the classes  \texttt{rdfs:Class} and \texttt{rdf:Property}.

\textbf{Computing Equivalence Set Graph.} 
As we saw in the previous section, for computing an ESG a preliminary step is needed in order to compute the deductive closure of properties (which is an ESG itself).
We can distinguish two cases depending if condition $\peq = \pe$ and $\psub = \ps$ holds or not.
If this condition holds (e.g. when the procedure is set for computing the ESG of properties), then for retrieving equivalences and specializations of $\peq$ and $\psub$ the procedure has to use the ESG is building (cf. \textsc{UpdatePSets}).
Otherwise, the procedure has to compute an ESG (i.e. $\cli{C}{\pe,\ps}{\pe,\ps}$) using $\pe$ as $\peq$ and  $\ps$ as $\psub$. 
We describe how the algorithm works in the first case (in the second case, the algorithm acts in a similar way, unless that $P_{e}$ and  $P_s$ are filled with $\cli{C}{\pe,\ps}{\pe,\ps}$($\peq$) and $\cli{C}{\pe,\ps}{\pe,\ps}$($\psub$) respectively and  \textsc{UpdatePSets} is not used).

The input of the main procedure (i.e. Algorithm~\ref{proc:main}) includes:
\begin{enumerate*}[label=\textit{(\roman*)}]
    
    \item a set $P_{e}$ of equivalence relations. 
    In our case $P_{e}$ will contain \texttt{owl:equivalentProperty} for the ESG of properties, and (the deductive closure of) \texttt{owl:equivalentClass} for the ESG of classes;
    
    \item a set $P_s$ of specialisation relations. 
    In our case $P_{s}$ will contain \texttt{rdfs:subPropertyOf} for the ESG of properties, and (the deductive closure of) \texttt{rdfs:subClassOf} for the ESG of classes.
    
\end{enumerate*}
The output of the algorithm is a set of maps and multi-maps which store nodes and edges of the computed ESG:
\begin{description}
	\item[\textit{ID}] a map that, given an IRI of an entity, returns the identifier of the ES it belongs to;
	
	\item[\textit{IS}] a multi-map that, given an identifier of an ES, returns the set of entities it contains;
	
	\item[\textit{H} ($H^-$)] a multi-map that, given an identifier of an ES, returns the identifiers of the explicit super (sub) ESs.
\end{description}
The algorithm also uses two additional data structures:
\begin{enumerate*}[label=\textit{(\roman*)}]
	
	\item $P'_{e}$ is a set that stores the  equivalence relations already processed (which are removed from $P_{e}$ as soon as they are processed);

	\item $P'_{s}$ is a set that stores the specialisations relations already processed (which are removed from $P_{s}$ as soon as they are processed).
	
\end{enumerate*}

The algorithm repeats three sub-procedures until $P_{e}$ and $P_s$ become empty: 
\begin{enumerate*}[label=\textit{(\roman*)}]

\item Compute Equivalence Sets (Algorithm~\ref{proc:equivalence}),    

\item  Compute the Specialisation Relation among the Equivalence Sets (Algorithm~\ref{proc:subOfIdentitySets}),

\item Update $P_{e}$ and $P_s$ (i.e. \textsc{UpdatePSets}).
\end{enumerate*}

Algorithm~\ref{proc:equivalence} iterates over $P_{e}$, and at each iteration moves a property $p$ from $P_{e}$ to $P'_{e}$, until $P_{e}$ is empty.
For each triple $\langle r_1,p,r_2 \rangle \in G$, it tests the following conditions and behaves accordingly:
\begin{enumerate}
    
    \item $r_1$ and $r_2$ do not belong to any ES, then: a new ES containing $\{r_1,r_2\}$ is created and assigned an identifier i. ($r_1$,i) and ($r_2$,i) are added to ID, and (i, $\{r_1,r_2\}$) to IS;
    
    \item $r_1$ ($r_2$) belongs to the ES with identifier $i_1$ ($i_2$) and $r_2$ ($r_1$) does not belong to any ES. Then ID and IS are updated to include $r_2$ ($r_1$) in $i_1$ ($i_2$);
    
    
    \item $r_1$ belongs to an ES with identifier $i_1$ and $r_2$ belongs to an ES with identifier $i_2$ (with $i_1\neq i_2$). Then $i_1$ and $i_2$ are merged into a new ES with identifier $i_3$ and the hierarchy is updated by Algorithm~\ref{proc:fixHierarchy}.
    This algorithm ensures both the followings: 
    \begin{enumerate*}[label=\textit{(\roman*)}]
    
    \item the super (sub) set of $i_3$ is the union of the super (sub) sets of $i_1$ and $i_2$;
    
    \item the super (sub) sets that are pointed by  (points to) (through $H$ or $H^-$)  $i_1$ or $i_2$, are pointed by (points to) $i_3$ and no longer by/to $i_1$ or $i_2$.
    
    \end{enumerate*}

\end{enumerate}

The procedure   for computing the specialization (i.e. Algorithm ~\ref{proc:subOfIdentitySets}) moves p from $P_s$ to $P'_s$ until $P_s$ becomes empty.
For each triple $\langle r_1, p, r_2 \rangle \in G$ the algorithm ensures that $r_1$ is in an equivalence set with identifier $i_1$ and $r_2$ is in an equivalence set with identifier $i_2$:
\begin{enumerate}
    \item If $r_1$ and $r_2$ do not belong to any ES, then
    IS and ID are updated to include two new ESs $\{r_1\}$ with identifier $i_1$ and $\{r_2\}$ with identifier $i_2$;
        
    \item if $r_1$ ($r_2$) belongs to an ES with identifier $i_1$ ($i_2$) and $r_2$ ($r_1$) does not belong to any ES, then 
     IS and ID are updated to include a new ES $\{r_2\}$ ($\{r_1\}$) with identifier $i_2$ ($i_1$).
    
    
\end{enumerate}
At this point $r_1$ is in $i_1$ and $r_2$ is in $i_2$ ($i_1$ and $i_2$ may be equal) and then $i_2$ is added to $H(i_1)$ and $i_1$ is added to $H^-(i_2)$. 

The procedure \textsc{UpdatePSets} (the last called by Algorithm~\ref{proc:main}) adds to $P_{e}$ ($P_{s}$)  the properties in the deductive closure of properties  in $P'_{e}$ ($P'_{s}$).
For each property p in $P'_{e}$ ($P'_s$), \textsc{UpdatePSets} uses ID to retrieve the identifier of the ES of p, then it uses $H^-$ to traverse the graph in order retrieve all the ESs that are subsumed by ID(p).
If a property $p'$ belongs to ID(p) or to any of the traversed ESs is not in $P'_{e}$ ($P_s$), then $p'$ is added to $P_{e}$ ($P_s$).


\textbf{Algorithm Time Complexity.} 
Assuming that retrieving all triples having a certain predicate and inserting/retrieving values from maps costs O(1). 
The algorithm steps once per each equivalence or subsumption triple. \textsc{FixHiearchy} costs in the worst case O($n_{eq}$) where $n_{eq}$ is the number of equivalence triples in the input dataset.  $n_{sub}$ is the number of specialization triples in the input dataset. 
Hence, time complexity of the algorithm is O($n_{eq}^2$ + $n_{sub}$).

\textbf{Algorithm Space Complexity.}
In the worst case the algorithm needs to create an equivalence set for each equivalence triple and a specialization relation for each specialization triple. 
Storing ID and IS maps costs $\sim$2n (where n is the number of observed entities from the input dataset), whereas storing H and $H^-$ costs $\sim4n^2$ .
Hence,  the space complexity of the algorithm is O($n^2$).

\begin{figure*}[ttt!]
\begin{minipage}[t]{0.4\textwidth}
  \vspace{0pt}  
  \begin{algorithm}[H]
	\caption{Main Procedure}\label{proc:main}
	\begin{algorithmic}[1]
		\Procedure{Main}{$P_{e},P_{s}$}
		
			\State $P^{'}_{e} = P^{'}_{s} = \emptyset$
	
			\State Init ID: $IRI \rightarrow$ $ID_{IS}$
			\State Init IS: $ID_{IS} \rightarrow IS$ 
			\State Init $H$: $ID_{IS} \rightarrow 2^{ID_{IS}}$
			\State Init $H^-$: $ID_{IS} \rightarrow 2^{ID_{IS}}$
			\State Init C: $ID_{IS} \rightarrow 2^{ID_{IS}}$
			\State Init $C^-$: $ID_{IS} \rightarrow 2^{ID_{IS}}$
			
			\While{$P_{e} \neq \emptyset || P_{s} \neq \emptyset$} \label{line:mainloop}
			
			\State \Call{ComputeESs}{ }
			\State \Call{ComputeHierarchy}{ }
			\State \Call{UpdatePSets}{ }
			\EndWhile
		\EndProcedure
	    \Procedure{UpdatePSets}{ }
	    \For{$p'_{e}\in P'_{e} || p'_{s} \in P'_{s}$}
	    \For{$p_{e}$ s.t.  $\cli{C}{p_e,p_s}{p_e,p_s}(p'_{e})  $}
	    \State Add $p_{e}$ to $P_{e}$ if $p_{e}\notin P'_{e}$
	    \EndFor
	    \For{$p_{s}$ s.t.  $\cli{C}{p_e,p_s}{p_e,p_s}(p'_{s})  $}
	    \State Add $p_{s}$ to $P_{s}$ if $p_{s}\notin P'_{s}$
	    \EndFor
	    \EndFor
	    \EndProcedure
	\end{algorithmic}
\end{algorithm}
\end{minipage} \hfill
\begin{minipage}[t]{0.6\textwidth}
  \vspace{0pt}
  \begin{algorithm}[H]
	\caption{Compute Equivalence Sets}\label{proc:equivalence}
	\begin{algorithmic}[1]
		\Procedure{ComputeESs}{ }
	
		\For{$p_e \in P_{e}$}
		\State Remove p from $P_{e}$ and Put p in $P^{'}_{e}$
		\For{$\langle r_1,p_e,r_2\rangle \in G$}
		\If {$ID(r_1) = \emptyset \land ID(r_2) = \emptyset $}
		\State Let i be a new identifier 
		\State Put $(r_1,i)$ and $(r_2,i)$  in ID 
		\State Put $(i,\{r_1,r_2\})$ in IS
		\ElsIf {$ID(r_1) = i_1 \land ID(r_2) = \emptyset $}
		\State Put $(r_2,i_1)$  in ID and Put $r_2$ in $IS(i_1)$
		\ElsIf {$ID(r_1) = \emptyset \land ID(r_2) = i_2 $}
		\State Put $(r_1,i_2)$  in ID and Put $r_1$ in $IS(i_2)$
		\ElsIf {${\scriptstyle  ID(r_1) = i_1 \land ID(r_2) = i_2 \land i_1 \neq i_2}$}
		\State Let $IS_3 \gets IS(i_1) \cup IS(i_1)$ 
		\State Let $i_3$ be a new identifier  
		\State Put $(i_3, IS_3)$ in IS
		\State Put $(r_3,i_3)$ in ID for all $r_3\in IS_3$
		\State Remove $(i_1,IS(i_1))$ from IS 
		\State Remove  $(i_2,IS(i_2))$ from IS 
		\State \Call{FixHierarchy}{$i_1$,$i_2$,$i_3$}
		\EndIf
		\EndFor
		\EndFor
		
		\EndProcedure
	\end{algorithmic}
\end{algorithm}
\end{minipage}
\hfill
\end{figure*}

\begin{figure*}[ttt!]
\begin{minipage}[t]{0.4\textwidth}
  \vspace{0pt}  
\begin{algorithm}[H]
	\caption{}
	\label{proc:fixHierarchy}
	\begin{algorithmic}[1]
		\Procedure{FixHierarchy}{$i_1,i_2,i_3$}
		\State $H(i_3) = H(i_1) \cup H(i_2)$
		\State $H^-(i_3) = H^-(i_1) \cup H^-(i_2)$
		\For{$i_{11}\in H(i_1)$}
		\State Remove $i_1$ from $H^-(i_{11})$ 
		\State Add $i_3$ to $H^-(i_{11})$
		\EndFor
		\For{$i_{11}\in H^-(i_1)$}
		\State Remove $i_1$ from $H(i_{11})$ 
		\State Add $i_3$ to $H(i_{11})$
		\EndFor
		\For{$i_{21}\in H(i_2)$}
		\State Remove $i_2$ from $H^-(i_{21})$
		\State Add $i_3$ to $H^-(i_{21})$
		\EndFor
		\For{$i_{21}\in H^-(i_2)$}
		\State Remove $i_2$ from $H(i_{21})$ 
		\State Add $i_3$ to $H(i_{21})$
		\EndFor
		\EndProcedure
	\end{algorithmic}
\end{algorithm}
\end{minipage} \hfill
\begin{minipage}[t]{0.6\textwidth}
\begin{algorithm}[H]
	\caption{}\label{proc:subOfIdentitySets}
	\begin{algorithmic}[1]
		\Procedure{ComputeHierarchy}{ }
		
		\For{$p_s \in P_{s}$}
		
		\State Remove p from $P_{s}$ and put p in $P^{'}_{s}$
		
		\For{$\langle r_1,p_s,r_2\rangle$}
	
		\If {$ID(r_1) = \emptyset \land ID(r_2) = \emptyset $}
		\State Let $i_1$ and $i_2$ be new identifiers
		\State Put $(r_1,i_1)$ and $(r_2,i_2)$  in ID
		\State  Put $(i_1,\{r_1\})$ and $(i_2,\{r_2\})$ in IS

		\ElsIf {$ID(r_1) = i_1 \land ID(r_2) = \emptyset $}
		\State Let  $i_2$ be a new identifier
		\State Put $(r_2,i_2)$  in ID
		and  $(i_2,\{r_2\})$ in IS
	
		\ElsIf {$ID(r_1) = \emptyset \land ID(r_2) = i_2 $}
		\State Let  $i_1$ be a new identifier
		\State Put $(r_1,i_1)$  in ID
		\State Put  $(i_1,\{r_1\})$ in IS
	
	
		\EndIf
		
			\State  Put $(i_1,H(i_1)\cup  \{i_2\})$ in H 
			\State Put $(i_2,H^-(i_2)\cup \{i_1\})$ in $H^-$
	
	\EndFor
	\EndFor
		
		\EndProcedure
	\end{algorithmic}
\end{algorithm}
\end{minipage}
\hfill
\end{figure*}

%% file: src/lodGraph.tex
In order to analyse the modeling structure and style of LOD we compute two ESGs from LOD-a-lot: one for classes and one for properties.
Both graphs are available for download\footnote{\url{https://w3id.org/edwin/iswc2019_esgs}}. We used a laptop (3Ghz Intel Core i7, 16GB of RAM). Building the two ESGs took $\sim$11 hours, computing their extension took $\sim$15 hours. Once the ESG are built, we can query them to compute the metrics defined in \ref{sec:metrics} and make observations at LOD scale within the order of a handful of seconds/minutes. Queries to compute indirect extensional dimension may take longer, in our experience up to 40 minutes.  

The choice of analysing classes and properties separately reflects the distinctions made by RDF(S) and OWL models.
However, this distinction is sometimes overlooked in LOD ontologies. We observed the presence of the following triples:
\begin{verbatim}
rdfs:subPropertyOf rdfs:domain rdf:Property .  # From W3C
rdfs:subClassOf rdfs:domain rdfs:Class .       # From W3C
rdfs:subClassOf rdfs:subPropertyOf rdfs:subPropertyOf . # From BTC
\end{verbatim}
The first two triples come from RDFS vocabulary defined by W3C, and the third can be found in the Billion Triple Challenge datasets\footnote{\url{https://github.com/timrdf/DataFAQs/wiki/Billion-Triples-Challenge}}.
These triples imply that if a property $p_1$ is subsumed by a property $p_2$, then $p_1$ and $p_2$ become classes.
Since our objective is to observe classes and property separately we can not accept the third statement.
For similar reasons, we can not accept the following triple:
\begin{verbatim}
rdf:type rdfs:subPropertyOf rdfs:subClassOf . # From BTC
\end{verbatim}
which implies that whatever has a type becomes a class.
It is worth noticing that these statements does not violate RDF(S) semantics, but they do have far-reaching consequences for the entire Semantic Web, most of which are unwanted.

\begin{table}[ht!]
\centering
    \begin{tabular}{l@{\hskip 0.2in}c@{\hskip 0.2in}c@{\hskip 0.2in}c}
        \hline
        \multicolumn{2}{c}{Metrics}
        &  Property & Class \\\hline
        
        \# of Observed Entities & $OE$ & 1,308,946  & 4,857,653 \\
        \# of Observed Entities without BNs & $OE_{bn}$  & 1,301,756  & 3,719,371  \\
        \# of Blank Nodes (BNs) & $BN$ & 7,190 & 1,013,224 \\
        \# of Equivalence Sets (ESs) & $ES$ & 1,305,364 & 4,038,722  \\ 
        \# of Equivalence Sets (ESs) without BNs  & $ES_{bn}$ & 1,298,174 & 3,092,523  \\ 
         Ratio between ES and OE & R &  .997 & .831\\
         Ratio between ES and OE without BNs & $R_{bn}$  &  .997 & .831\\
        \# of Edges  & E & 147,606 & 5,090,482 \\
        Maximum Height & $H_{max}$ & 14 & 77 \\
        \# Isolated ESs & $IN$ & 1,157,825 & 288,614 \\
        \# of  Top Level ESs & $TL$ & 1,181,583 & 1,281,758 \\
        \# of  Top Level ESs without BNs & $TL_{bn}$  & 1,174,717 & 341,792 \\
        \# of OE in Top Level ESs & $OE$-$TL$ & 1,185,591& 1,334,631\\
        \# of OE in Top Level ESs without BNs & $OE$-$TL_{bn}$ & 1,178,725 & 348,599 \\
        Ratio between TL and OE-TL & $RTL$ & .996 & .960 \\
        Ratio between TL and OE-TL without BNs & $RTL_{bn}$  & .996 & .980 \\
        \# of Weakly Connected Components & $WCC$ & 1,174,152 & 449,332 \\
        \# of Strongly Connected Components & $SCC$ & 1,305,364 & 4,038,011 \\
        \# of OE with Empty Extension & $OE_{0}$ & 140,014 & 4,024,374 \\
        \# of OE with Empty Extension without BNs & $OE_{0bn}$ & 132,824  & 2,912,700 \\
        \# of ES with Empty Extension & $ES_{0}$ & 131,854 & 3,060,467 \\
        \# of ES with Empty Extension without BNs &  $ES_{0bn}$  & 124,717 & 2,251,626  \\
        \# of ES with extensional size greater than 1 & $IES(1)$ & 1,173,510 & 978,255 \\
        \# of ES with extensional size greater than 10 &  $IES(10)$ & 558,864  & 478,746 \\
        \# of ES with extensional  size greater than 100 & $IES(100)$ & 246,719 & 138,803 \\
        \# of ES with extensional  size greater than 1K & $IES(1K)$ & 79,473 & 30,623 \\
        \# of ES with extensional size greater than 1M & $IES(1M)$ & 1,762 & 3,869 \\
        \# of ES with extensional  size greater than 1B  & $IES(1B)$ & 34  &  1,833 \\
        \# of OE-TL with Empty Extension & $OE$-$TL_{0}$ & 26,640 & 1,043,099 \\
        \# of OE-TL with Empty Extension w/o BNs & $OE$-$TL_{0bn}$ & 19,774 & 83,674\\
        \# of TL with Empty Extension & $TL_{0}$ & 18,884 & 869,443 \\
        \# of TL with Empty Extension w/o BNs &  $TL_{0bn}$ & 12,071 & 66,805  \\
        \hline
        
    \end{tabular}
   
    \caption{Statistics computed on the equivalent set graph for properties and classes, from  LOD-a-lot. They include the metrics defined in Section~\ref{sec:metrics}. IES(n) indicates the Number of Equivalent Sets having indirect size n or greater. The term \emph{entity} is here used to refer to classes and properties.}
    \label{tab:ES_graph_properties_classes}
\end{table}

\noindent\textbf{Equivalence Set Graph for Properties.}
We implemented the algorithm presented in Section~\ref{sec:algorithm} to compute the ESG for properties contained in LOD-a-lot~\cite{shortFernandez2017}.
Our input parameters to the algorithm are:
\begin{enumerate*}[label=\textit{(\roman*)}]

    \item $P_{eq}$ $=$ \{\texttt{owl:equivalentProperty}\};
    
    \item $P_{s}$ $=$ \{\texttt{rdfs:subPropertyOf}\}.
    
\end{enumerate*}
Since \texttt{owl:equivalentProperty} is neither equivalent to nor subsumed by any other property in LOD-a-lot, the algorithm used only this property for retrieving equivalence relations.
Instead, for computing the hierarchy of equivalence sets the algorithm used 451 properties which have been found implicitly equivalent to or subsumed by \texttt{rdfs:subPropertyOf}. 

\begin{figure}[t]
  \centering
  \begin{subfigure}[t]{0.48\textwidth}
    \includegraphics[width=\textwidth]{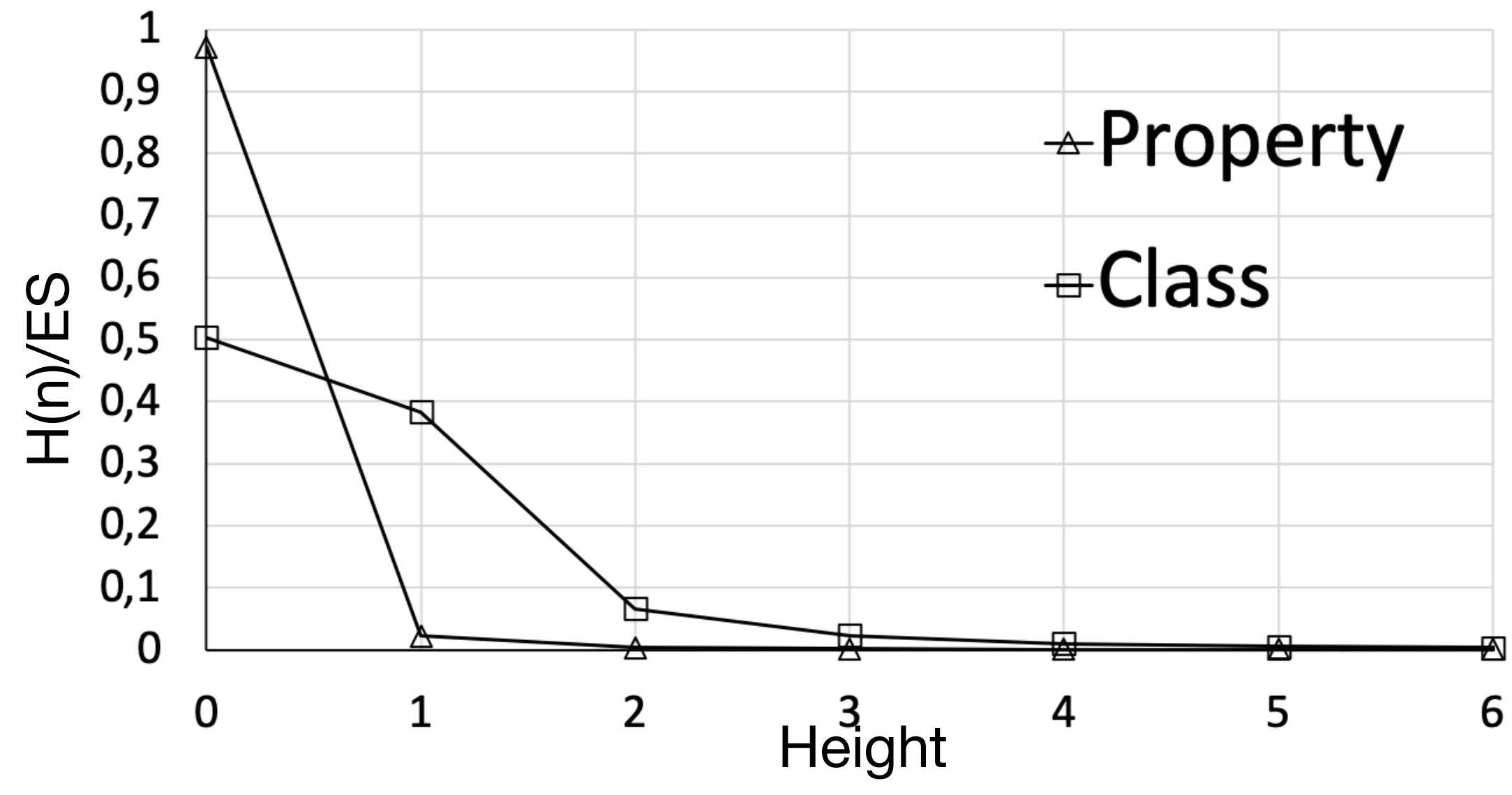}
    \caption{Height distribution of ES}
    \label{fig:height_distribution}
  \end{subfigure}
  ~%
  \begin{subfigure}[t]{0.48\textwidth}
    \includegraphics[width=\textwidth]{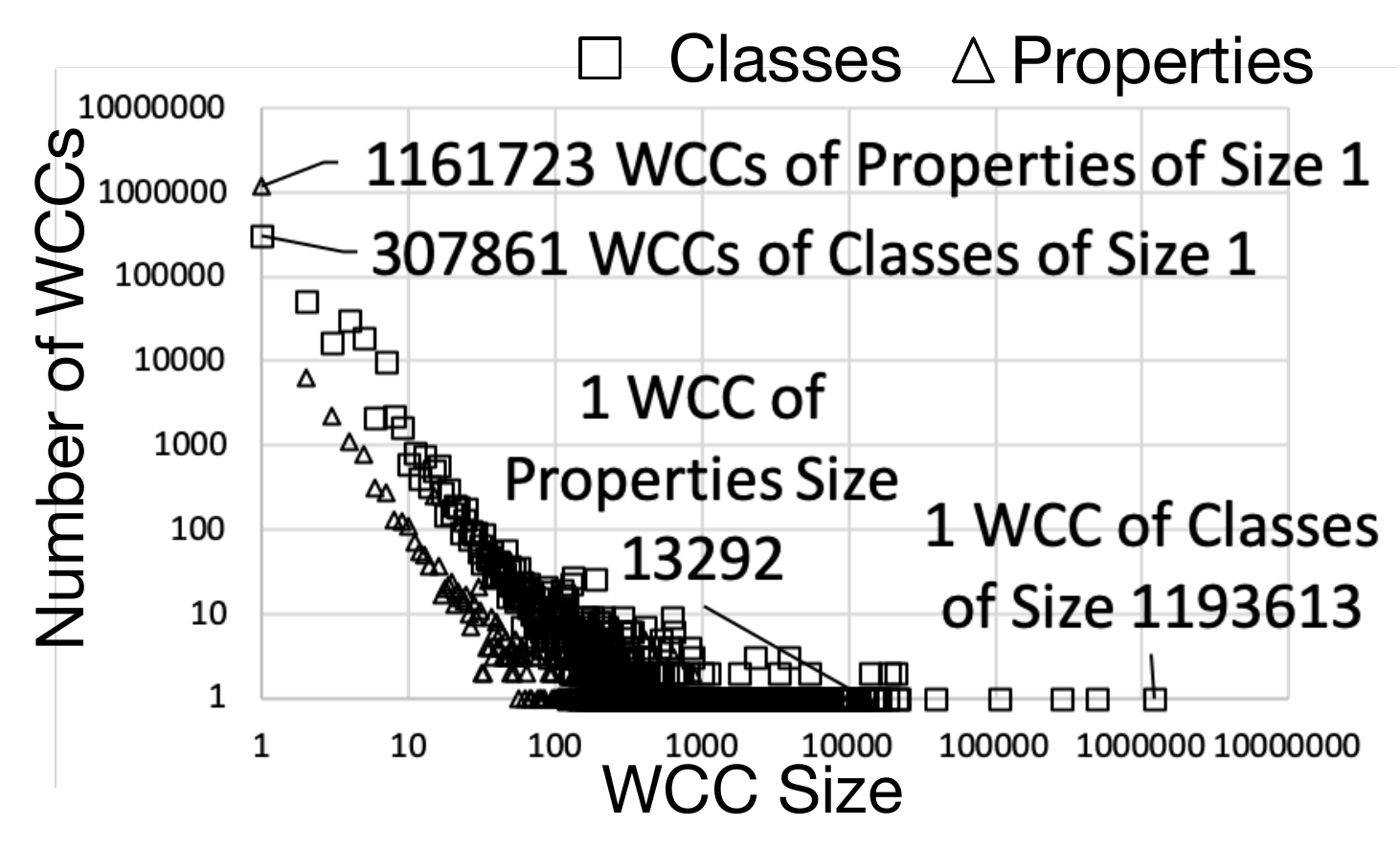}
    \caption{Distribution of the size of Weakly Connected Components. The function shows how many WCC have a certain size.}
    \label{fig:wcc_distribution}
  \end{subfigure}
  \begin{subfigure}[t]{0.48\textwidth}
    \includegraphics[width=\textwidth]{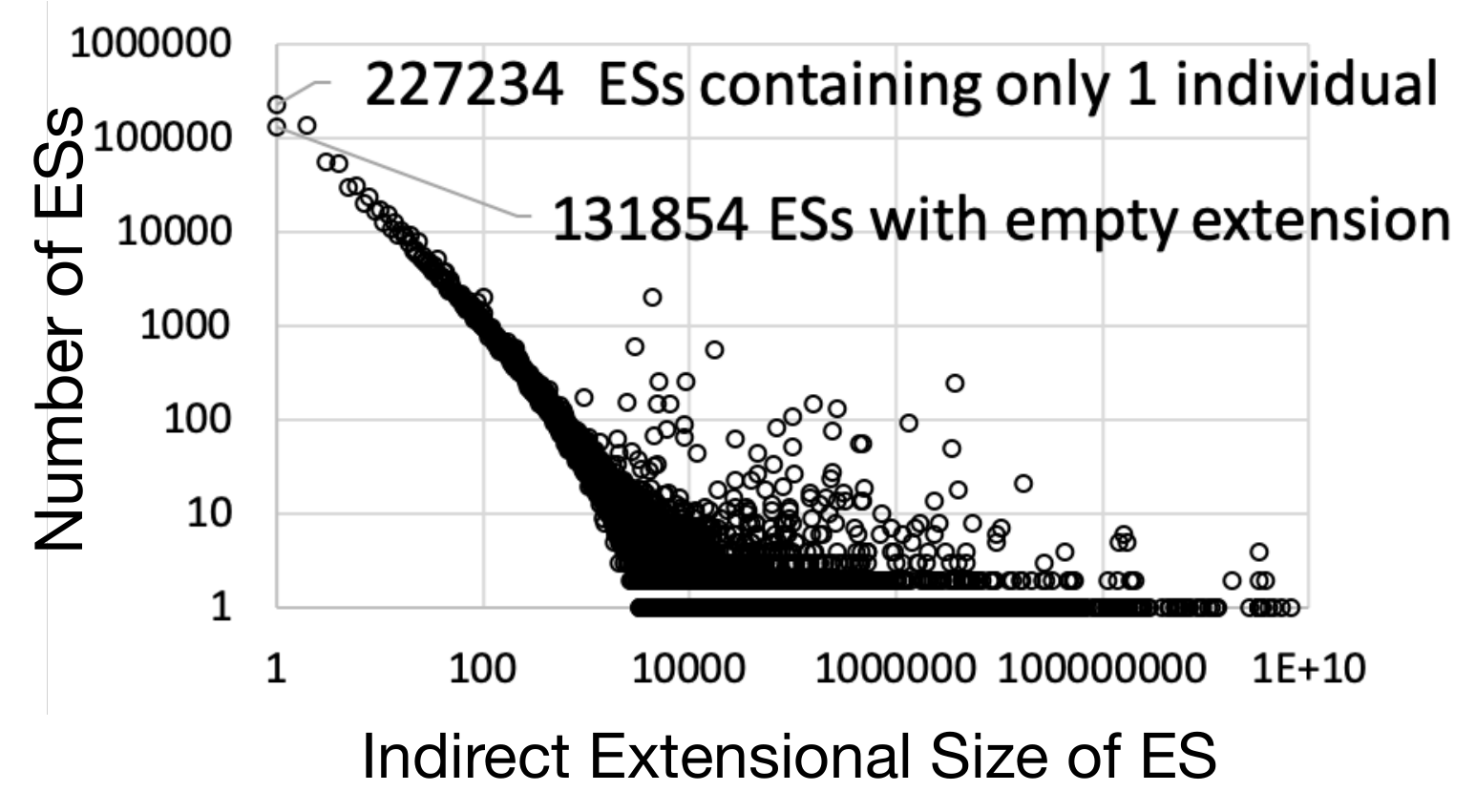}
    \caption{Distribution of IES for properties: the extension size of property ES. The function indicates how many ES have a certain extension size.}
    \label{fig:desi_properties}
  \end{subfigure}
  ~%
  \begin{subfigure}[t]{0.48\textwidth}
    \includegraphics[width=\textwidth]{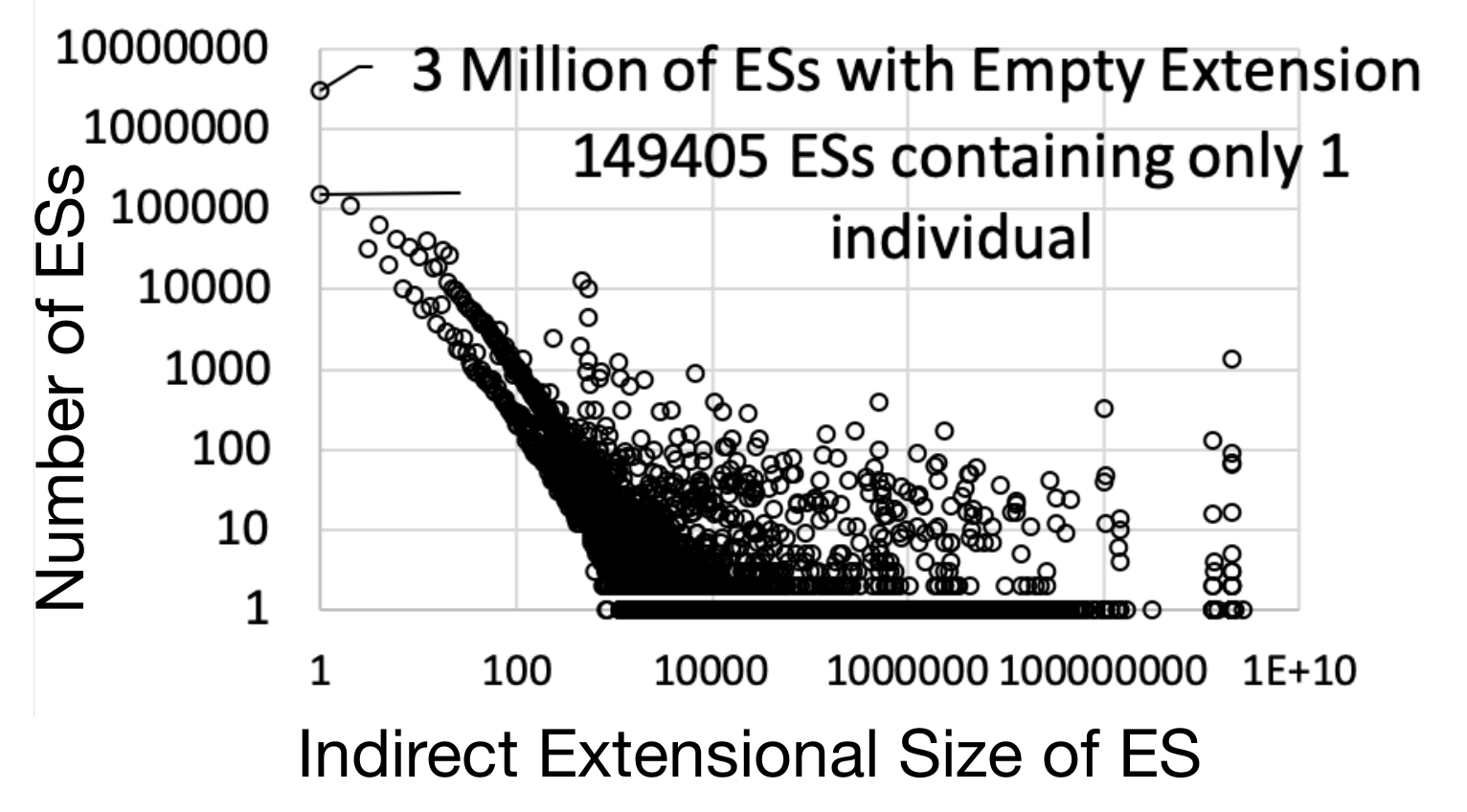}
    \caption{Distribution of IES for classes: the extension size of class ES. The function indicates how many ES have a certain extension size.}
    \label{fig:desi_classes}
  \end{subfigure}
  \label{fig:stats}
  \caption{Figure~\ref{fig:height_distribution} shows the normalised number of nodes per  height, Figure~\ref{fig:wcc_distribution} shows the number of weakly connected component per component size, Figure~\ref{fig:desi_properties} and Figure~\ref{fig:desi_classes} show the number of ESs per indirect extensional size.
  Figures ~\ref{fig:wcc_distribution}, \ref{fig:desi_properties} and  \ref{fig:desi_classes} are in logarithmic scale.}
\end{figure}

Table~\ref{tab:ES_graph_properties_classes} presents the metrics (cf. Section~\ref{sec:metrics}) computed from the equivalence set graph for properties.
It is quite evident that the properties are poorly linked.
\begin{enumerate*}[label=\textit{(\roman*)}]
    
    \item The ratio (R) tends to 1, indicating that few properties are declared equivalent to other properties;
    
    \item the ratio between the number of equivalence sets (ES) and the number of isolated sets (IN) is 0.88, indicating that most of properties are defined outside of a hierarchy;
    
    \item the height distribution of ESG nodes (cf. Figure~\ref{fig:height_distribution}) shows that all the nodes have height less than 1;
    
    \item the high number of Weakly Connected Components (WCC) is close to the total number of ES.
    
\end{enumerate*}
Figure~\ref{fig:desi_properties} shows that the dimension of ESs follows the Zipf's law (a trend also  observed in~\cite{shortDing2005}): many ESs with few instances and few ESs with many instances.
Most properties ($\sim$90\%) have at least one instance. This result is in contrast with one of the findings of Ding and Finin in 2006~\cite{shortDing2006} who observed that most properties have never been instantiated. We note that blank nodes are present in property hierarchies, although they cannot be instantiated. This is probably due to some erroneous statement.

\textbf{Equivalence Set Graph for Classes.}
From the ESG for properties we extract all the properties implicitly equivalent to or subsumed by \texttt{owl:equivalentClass} (2 properties) and put them in $P_{eq}$, the input parameter of the algorithm. $P_s$ includes 381 properties that are implicitly equivalent to or subsumed by \texttt{rdfs:subClassOf}.

Table~\ref{tab:ES_graph_properties_classes} reports the metrics (cf. Section~\ref{sec:metrics}) computed from the ESG for classes.
Although class equivalence is more common than property equivalence, the value of R is still very high (0.83), suggesting that equivalence relations among classes are poorly used. Differently from properties, classes form deeper hierarchies: the maximum height of a node is 77 (compared to 14 for properties), only 7\% of nodes are isolated and only 31\% are top level nodes, we observe from Figure~\ref{fig:height_distribution} that the height distribution has a smoother trend than for properties but still it quickly reaches values slightly higher than 0.
We observe that (unlike properties) most of class ES are not instantiated: only 31.7\% of ES have at least one instance. A similar result emerges from the analysis carried out in 2006 by Ding and Finin~\cite{shortDing2006} who reported that 95\% of semantic web terms (properties and classes) have no instances (note that in~\cite{shortDing2006} no RDFS and OWL inferencing was done).
It is worth noticing that part (800K) of these empty sets contain only black node that cannot be directly instantiated. As for properties, the dimension of ES follows the Zipf's distribution (cf. Figure~\ref{fig:desi_classes}), a trend already observed in the early stages of the Semantic Web~\cite{shortDing2006}.
We also note that blank nodes are more frequent in class hierarchies than in property hierarchies (25\% of ES of classes contain at least one blank node).

%% file: src/discussion.tex
We have presented  an  empirical  study  aiming  at  understanding the modeling style and the overall semantic structure of the Linked Open Data cloud. We observed how classes, properties and individuals are used in  practice, and  we  also  investigated  how  hierarchies  of  concepts  are  structured, and how much they are linked. 

Even if our conclusions on the issues with LOD data are not revolutionarily (the community is in general aware of the stated problems for Linked Data), we have presented a framework and concrete metrics to obtain concrete results that underpin these shared informal intuitions. We now briefly revisit our main findings:

\textbf{LOD ontologies are sparsely interlinked.}
The values computed for metric R (ratio between ES and OE) tell us that LOD classes and properties are sparsely linked with equivalence relations. 
We can only speculate as to whether ontology linking is considered less important or more difficult than linking individuals, or wether the unlinked classes belong to very diverse domains.  However, we find a high value for metric TL (top level ES) with an average of $\sim$ 1.1 classes per ES. Considering that the number of top level classes (without counting BN) is $\sim$348k, it is reasonable to suspect a high number of conceptual duplicates. The situation for properties is even worse: the average number of properties per TL ES is 1 and the number of top level properties approximates their total number.

LOD ontologies are also linked by means of specialisation relations (\texttt{rdfs:subClassOf} and \texttt{rdfs:subPropertyOf}). Although the situation is less dramatic here, it confirms the previous finding.  As for properties, $\sim$88.7\% of ES are isolated (cf. IN).  Classes exhibit \emph{better} behaviour in this regard, with only 7\% of isolated classes. This confirms that classes are more linked than properties, although mostly by means of specialisation relations.  

\textbf{LOD ontologies are mostly flat.}
The maximum height of ESG nodes is 14 for properties and 77 for classes. Their height's distribution (Figure~\ref{fig:height_distribution}) shows that almost all ES ($\sim$ 100\%) belong to flat hierarchies. This observation, combined with the values previously observed (cf. IN and R), reinforces the claim that LOD must contain a large number of duplicate concepts.

As for classes, $\sim$50\% of ES have no specialising concepts, i.e., height=0 (Figure \ref{fig:height_distribution}). However, a bit less than the remaining ES have at least one specialising ES. Only a handful of ES reach up to 3 hierarchical levels. 
The WCC distribution (Figure~\ref{fig:wcc_distribution}) confirms that classes in non-flat hierarchies are mostly organised as siblings in short-depth trees.
We speculate that ontology engineers put more care into designing their classes than they put in designing their properties.

\textbf{LOD ontologies contain many uninstantiated concepts.}  
We find that properties are mostly instantiated ($\sim$90\%), which suggests that they are defined in response to actual need. However, most classes -- even not counting blank nodes -- have no instances: $\sim$ 67\% of TL ES have no instances. A possible interpretation is that ontology designers tend to over-engineer ontologies beyond their actual requirements, with overly general concepts. 

\subsection{Future work}


We are working on additional metrics that can be computed on ESGs, and on extending the framework to analyse other kinds of relations (e.g. disjointness). We are also making a step towards assessing possible relations between the domain of knowledge addressed by LOD ontologies and the observations made.